\documentclass[sn-mathphys]{sn-jnl}

\usepackage{natbib}
\usepackage{enumerate}
\usepackage{lmodern}
\usepackage{pdfpages}

\jyear{2023}%

\begin{document}

\title[]{Severe flooding and cause-specific hospitalization in the United States}


\author*[1]{\fnm{Sarika} \sur{Aggarwal}} 

\author[1]{\fnm{Jie} K. \sur{Hu}}

\author[2]{\fnm{Jonathan} A. \sur{Sullivan}}

\author[3]{\fnm{Robbie} M. \sur{Parks}}

\author[1]{\fnm{Rachel} C. \sur{Nethery}}

\affil[1]{\small \orgdiv{Department of Biostatistics}, \orgname{Harvard T.H. Chan School of Public Health} \\ \orgaddress{\street{677 Huntington Avenue}, \city{Boston}, \state{MA}, \country{United States}}}

\affil[2]{\small \orgdiv{School of Geography, Development, \& Environment}, \orgname{University of Arizona} \\ \orgaddress{\street{1064 E. Lowell Street}, \city{Tucscon}, \state{AZ}, \country{United States}}}

\affil[3]{\small \orgdiv{Department of Environmental Health Sciences}, \orgname{Columbia University} \\ \orgaddress{\street{722 West 168th Street}, \city{New York}, \state{NY}, \country{United States}}}


\abstract{Flooding is one of the most disruptive and costliest climate-related disasters and presents an escalating threat to population health due to climate change and urbanization patterns. 
Previous studies have investigated the consequences of flood exposures on only a handful of health outcomes and focus on a single flood event or affected region. 
To address this gap, we conducted a nationwide, multi-decade analysis of the impacts of severe floods on a wide range of health outcomes in the United States by linking a novel satellite-based high-resolution flood exposure database with Medicare cause-specific hospitalization records over the period 2000-2016. Using a self-matched study design with a distributed lag model, we examined how cause-specific hospitalization rates deviate from expected rates during and up to four weeks after severe flood exposure. 
Our results revealed that risk of hospitalization was consistently elevated during and for at least four weeks following severe flood exposure for nervous system diseases (3.5 \%; 95 \% confidence interval [CI]: 0.6 \%, 6.4 \%), skin and subcutaneous tissue diseases (3.4 \%; 95 \% CI: 0.3 \%, 6.7 \%), and injury and poisoning (1.5 \%; 95 \% CI: -0.07 \%, 3.2 \%). Increases in hospitalization rate for these causes, musculoskeletal system diseases, and mental health-related impacts varied based on proportion of Black residents in each ZIP Code. 
Our findings demonstrate the need for targeted preparedness strategies for hospital personnel before, during, and after severe flooding.}




\maketitle


\section{Introduction}\label{sec1}

Since 2000, flooding is the most frequently occurring climate-related hazard \citep{human_cost_2000, du_health_2010}, and increases in frequency and intensity are projected in the coming decades due to climate change, urbanization, and increasing settlement in floodplains  \citep{esd_trends_2018,alfieri_warmer_2017,ceola_floodplains_2014}. In the United States, floods cause over $30$ billion dollars of damage annually, yet funds for disaster relief, recovery and preparedness are dwindling \citep{wing_inequitable_2022, first_street_foundation}. Floods have deleterious economic and social consequences including human displacement, mortality, property damage, and environmental degradation and destruction\citep{mallakpour_changing_2015,rufat_vulnerability_2015,committee_on_urban_flooding_in_the_united_states_framing_2019, tellman2020using}. Investigations on flood damage and loss have typically been limited to a handful of health outcomes and localized case studies \citep{alderman2012floods,saulnier2017no,zhong2018long}. The lack of observational evidence on a wide array of flood-related health outcomes hinders reporting on global disaster risk reduction targets and our ability to assess successful adaptation against benchmarks \citep{sendai}. This knowledge gap impedes informed strategic preparedness and resilience-building efforts that could minimize the adverse health impacts of flood disasters, now and in the future.
\\ \\
Previous studies have identified a few areas of flood-related health impacts \citep{du_health_2010} including injuries \citep{ahern_global_2005,sindall_drowning_2022}, gastrointestinal and skin infections (primarily but not exclusively in low-income countries) \cite{alderman2012floods,chanda2010impact,reacher2004health,schwartz2006diarrheal}, chronic cardiovascular illness \citep{lowe_factors_2013}, and increased risk of mental health problems \citep{fernandez_mental_2015} following flood exposures. 
However, beyond these causes, there are numerous plausible pathways through which floods may impact health much more broadly. First, evacuations, displacement, and medical facility closures may lead to disruption in medication usage and disease treatment. In addition, stress brought about by storms and floods has been shown to be associated with health-degrading behaviors such as decreased physical activity and weight gain\citep{bell2019health}, smoking relapse among former smokers\citep{lanctot2008effects}, and increased alcohol use and abuse\citep{cerda2011prospective}. Further, exposure to dampness/mold and potentially contaminated floodwaters could cause a wide range of illnesses.
\\ \\
In the United States, flood impacts are not borne equally across groups. Numerous city-specific \cite{selsor_inequity_2023,debbage_injustice_2019} and nationwide \cite{wing_inequitable_2022,qiang_disparities_2019} studies have found that socially vulnerable populations, such as racial/ethnic minorities, low-income groups, and older adults, disproportionately reside in high flood-risk zones. Furthermore, projections suggest that Black communities will face the most severe increases in flood risk due to anthropogenic change in the coming decades \citep{wing_inequitable_2022}. 
\\ \\
Therefore, there is a critical need to evaluate and quantify how flood events over large spatio-temporal settings are associated with a broader spectrum of health outcomes to better anticipate and mitigate the full scope of health burdens of future floods by identifying vulnerable communities.  
\\ \\
In this work, our aim was to examine how 13 diverse categories of cause-specific hospitalization are associated with flood exposure in older adults during a 17-year study period for each affected ZIP Code in the United States. Our comprehensive investigation combines population-level cause-specific hospitalization Medicare records and high-resolution satellite-based flood exposure measures for over 70 major flood events occurring across the contiguous United States throughout nearly two decades \citep{cms, tellman_satellite_2021}. Using rigorous design and analytic strategies to isolate flood effects, we investigated if and how the risk of each type of health outcome changes relative to a typical day during and for one month following exposure, capturing both direct and indirect health impacts in the immediate and medium-term aftermath of flood exposures. Moreover, we investigated disproportionate health burdens arising from flooding between communities with larger Black populations and those without. We observed (i) increased hospitalization rates for broad disease causes during and throughout the four weeks following severe flood exposure; (ii) flood exposure of extreme severity heightened impacts on cause-specific hospitalization rates in comparison to moderate severity flood events; (iii) and cause-specific hospitalization rates increase for both non-Black and Black communities through different mechanisms during flood exposure.  
Our findings can inform development of flood prevention, mitigation, and management strategies to minimize health burdens in the United States and other high-income countries.

\section{Results}\label{sec2}

\subsection{Flood exposure}\label{sec2_exposure}
\begin{figure}
    \centering
    \includegraphics[scale = 0.45]{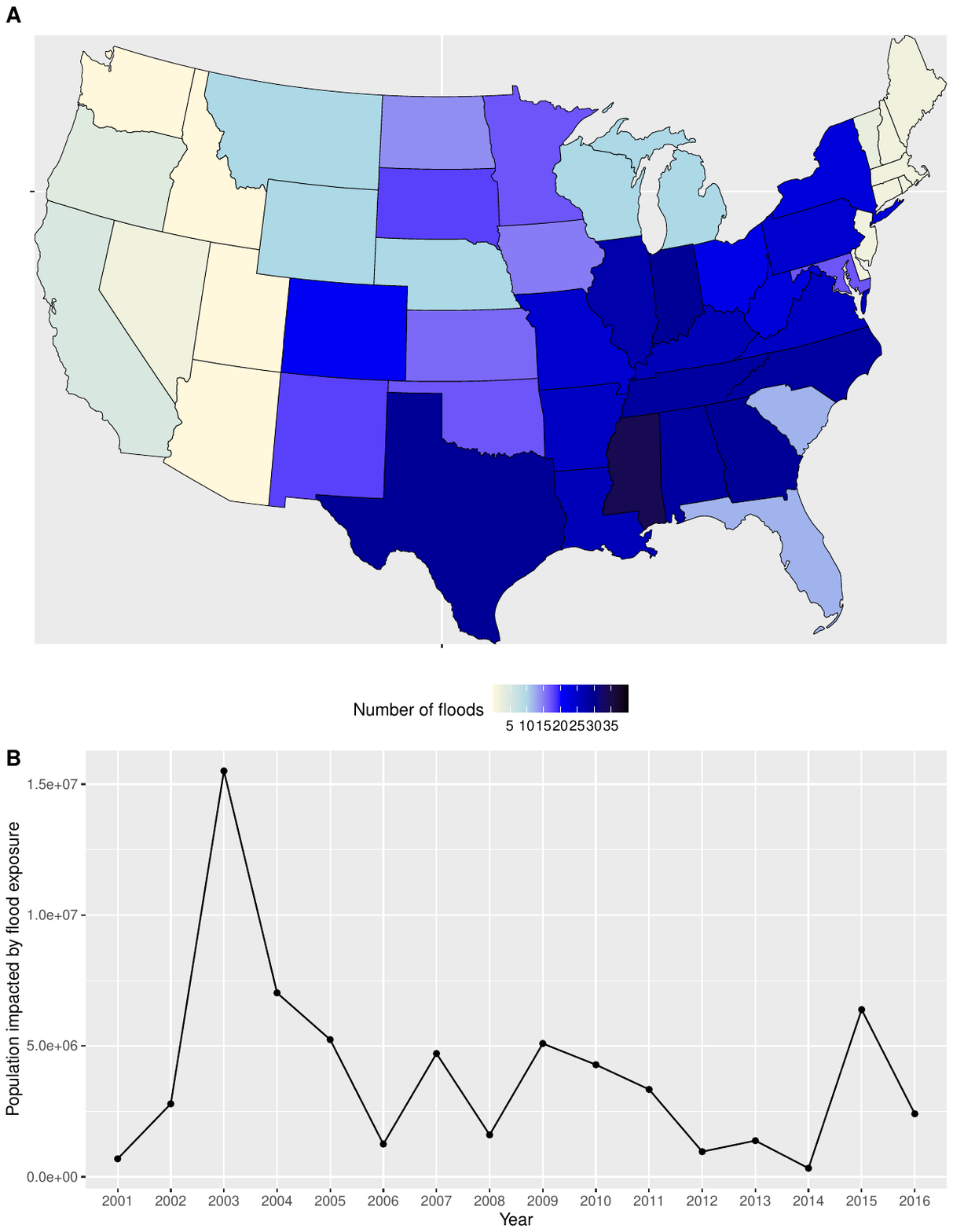}
    \caption{Spatial (A) and temporal (B) trends in flood exposure across the United States in 2000-2016}
    \label{spatio-temporal trends}
\end{figure}

Nationally, there were 75 flood events that ranged in duration from 1 day to 42 days, for a total of 353,925 flooded ZIP Code-days between 2000 and 2016. These flood events covered 11,594 unique ZIP Codes in 46 states and the District of Columbia (Figure \ref{spatio-temporal trends}A). 
Severe flood events were most common in the Ohio River Valley, the Southeast, and the Mid-Atlantic states. For example, in ZIP Code 37725, located in Dandridge, Tennessee (Jefferson County), there were 23 severe flood exposures during our study period, and was flooded for a total of 212 days. Furthermore, 81.3 \% of flood events were caused by heavy rain with the remaining 18.7 \% a result of tropical storms and surges. Each flood in the Global Flood Database is classified by severity (in accordance with the Dartmouth Flood Observatory classification) that ranks severity as 1, 1.5, or 2, and we refer to these severity categorizations as ``moderate'', ``high'' and ``extreme'', respectively. In our study, 40 out of 75 flood events were classified as moderate, 10 as high, and 25 as extreme. 

\subsection{Medicare hospitalizations}

\begin{figure}
    \centering
    \includegraphics[scale = 0.42]{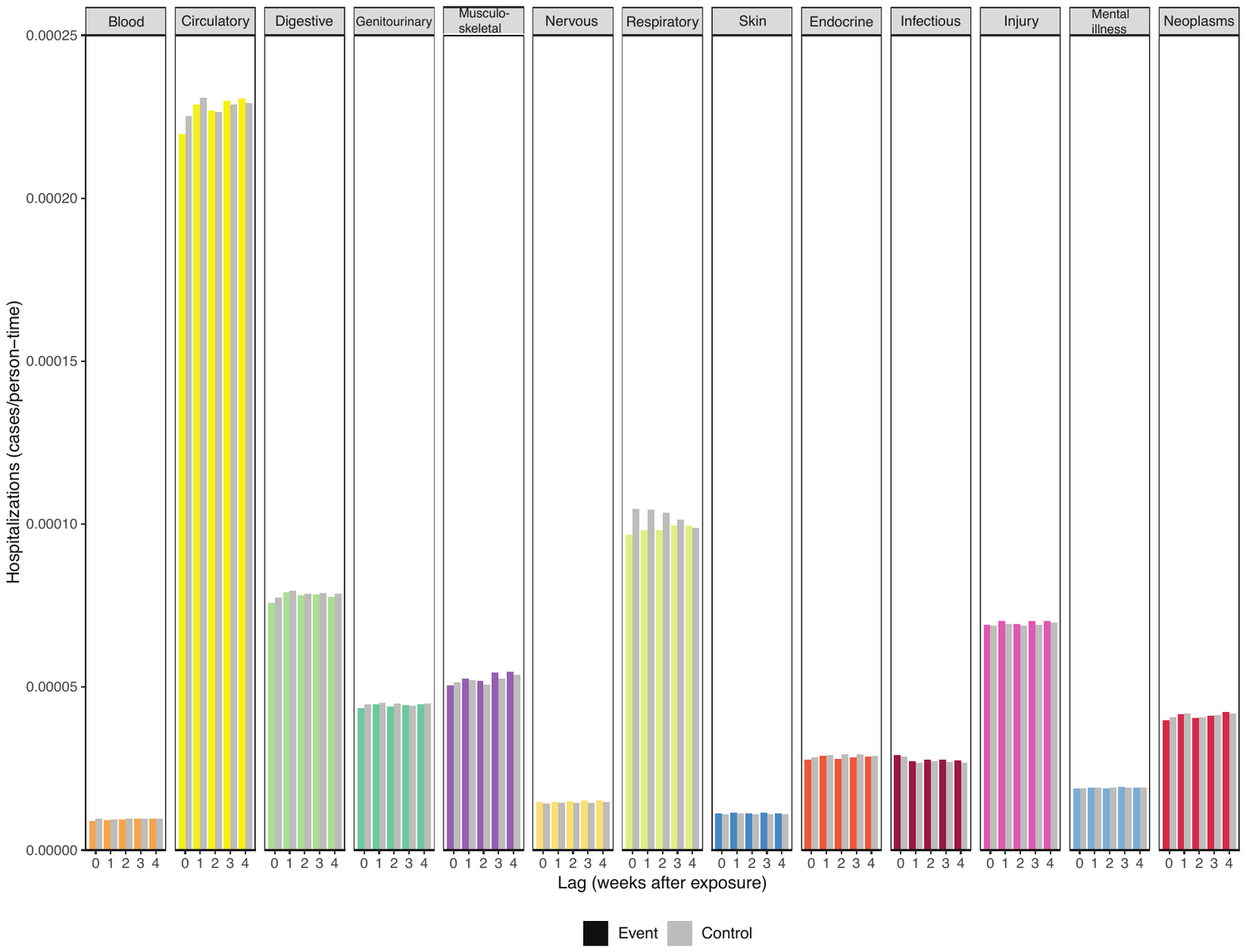}
    \caption{Rates of hospitalization by cause and lag time for ZIP Codes with at least one flood exposure in 2000-2016}
    \label{hospitalization_rates}
\end{figure}

From 2000 to 2016, there were 4,879,643 recorded hospitalizations among Medicare enrollees residing in 11,594 ZIP Codes that experienced at least one severe flood event. We grouped hospitalizations by primary cause using the Clinical Classifications Software algorithm into 13 broad causes of hospitalization that are well-defined and pertinent to the dynamic Medicare cohort we consider \citep{elixhauser2014clinical}. Rates of hospitalization by cause for events and corresponding controls during the flood exposure period (lag 0) and in each of the four weeks following exposure (lag weeks 1-4) are shown in Figure \ref{hospitalization_rates}. 
Diseases of the circulatory system (31 \%), respiratory system (13.9 \%), digestive system (10.8 \%), and injury and poisoning (9.6\%) were the four leading causes of hospitalization in our analytical sample. During flood exposure, we observed higher rates of hospitalization in controls for chronic conditions and in events for acute causes.



\subsection{Flood-attributable changes in hospitalization rates}
\begin{figure}
    \centering
    \includegraphics[scale = 0.42]{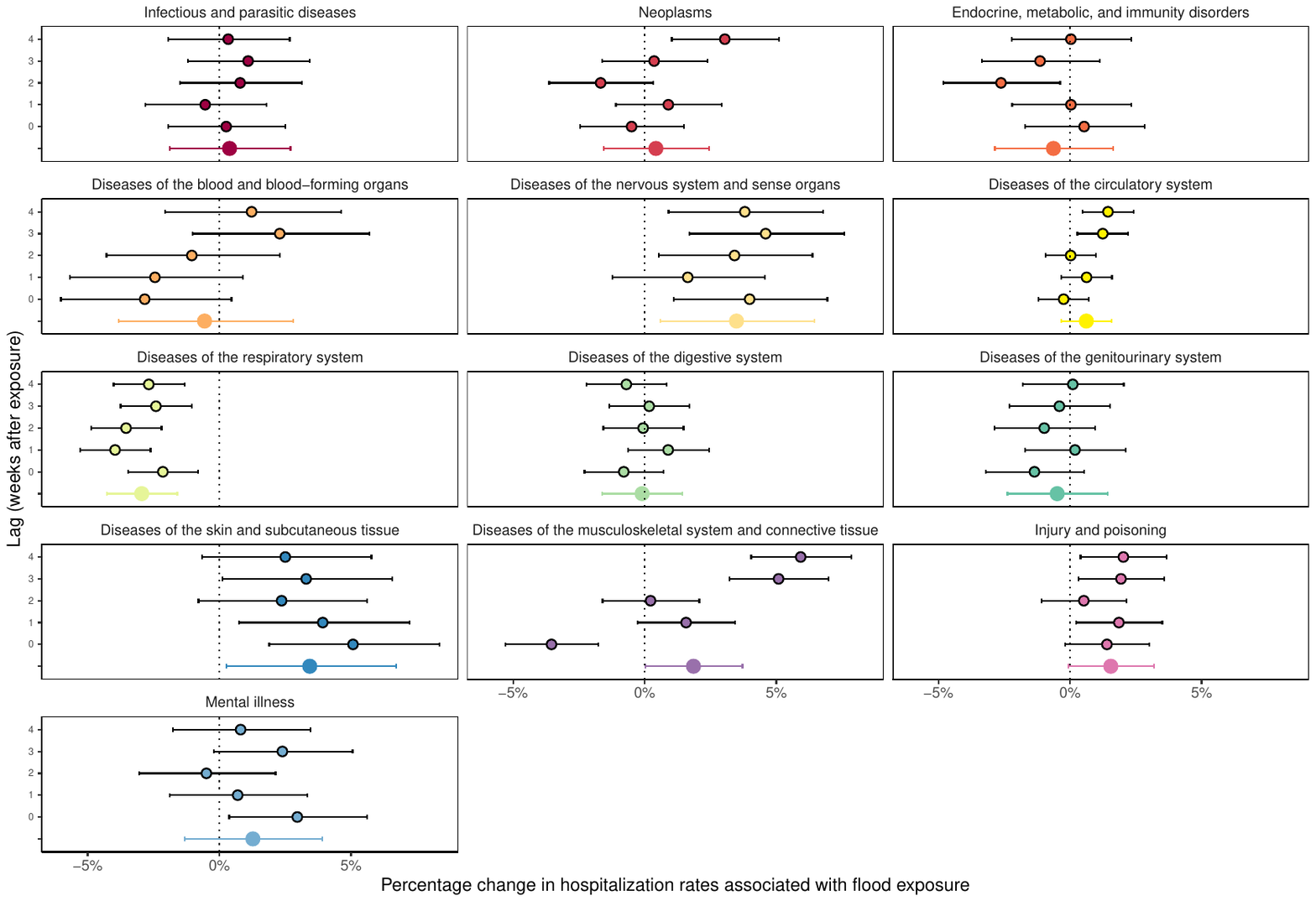}
    \caption{\textbf{Percent change in cause-specific hospitalization rates with flood exposure by cause and lag time.} Solid color point estimates (with corresponding Bonferroni-corrected 95 \% confidence intervals) represent average flood-attributable percent changes in hospitalization over the exposure period and the four lag weeks.}
    \label{level1_results}
\end{figure}

\begin{figure}
    \centering
    \includegraphics[scale = 0.42]{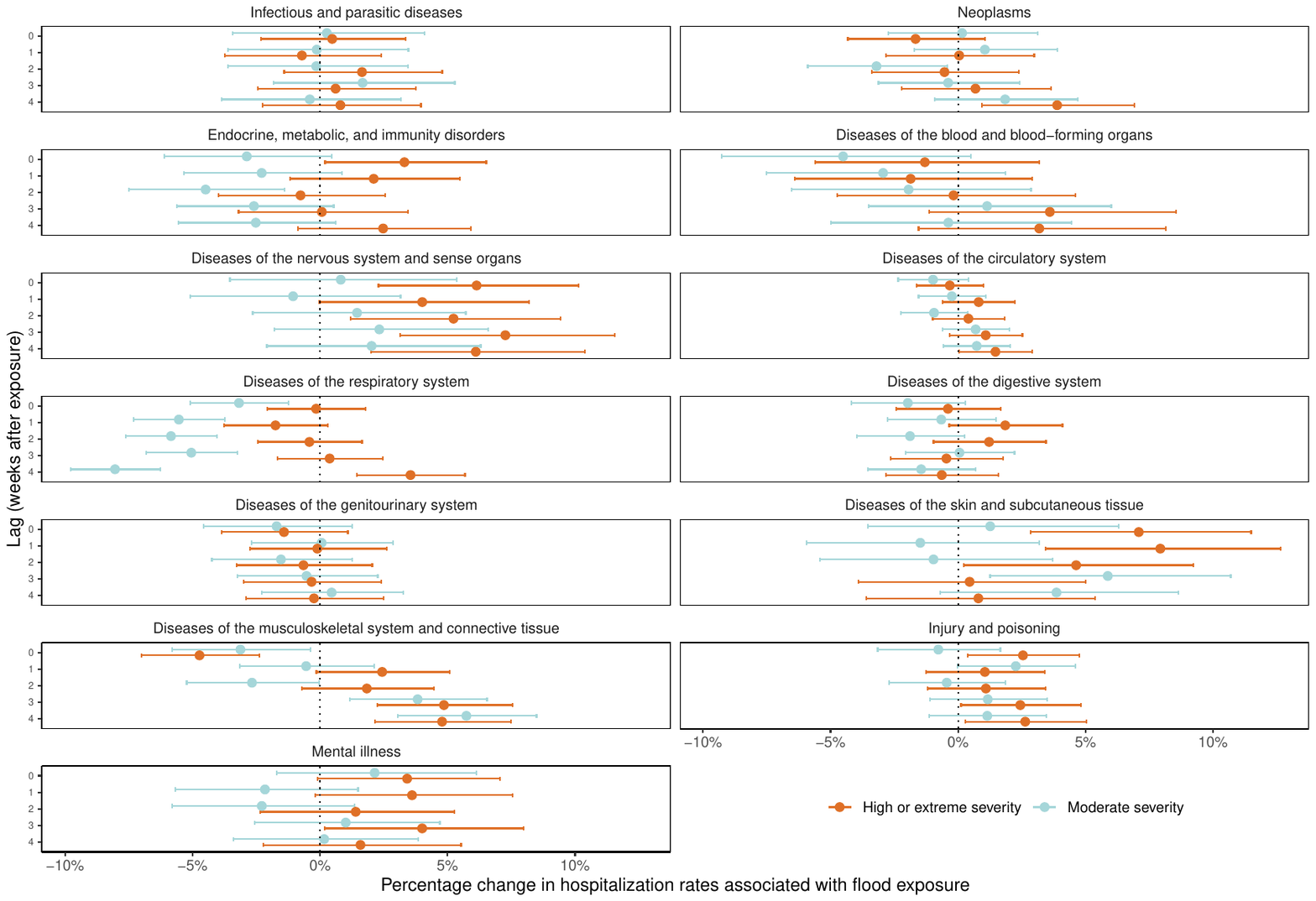}
    \caption{Percent change in cause-specific hospitalization rates with flood exposure by cause, severity, and lag time}
    \label{severity_results}
\end{figure}

\begin{figure}
    \centering
    \includegraphics[scale = 0.42]{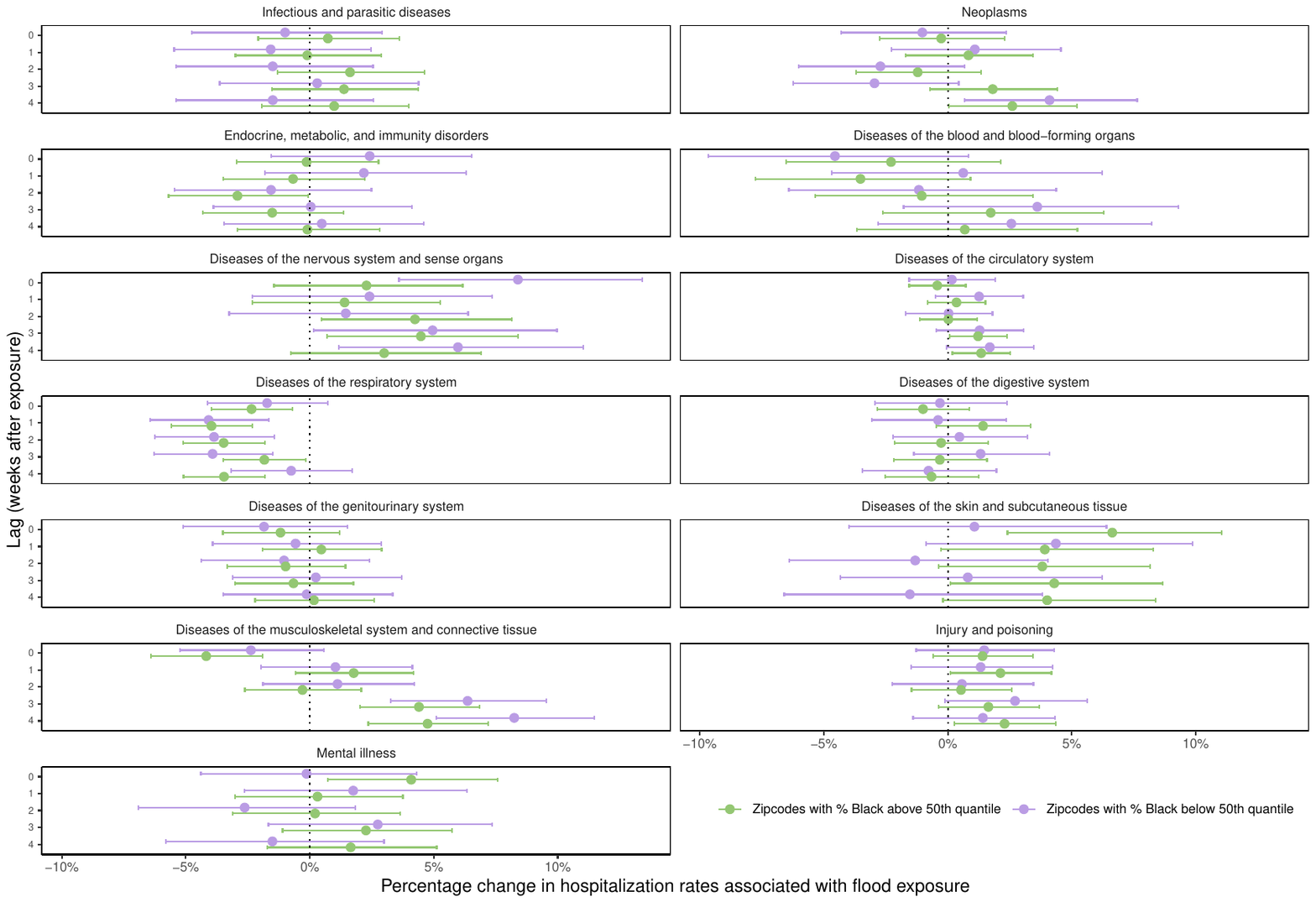}
    \caption{Percent change in cause-specific hospitalization rates with flood exposure by cause, proportion of Black residents, and lag time}
    \label{blk_results}
\end{figure}

We report the estimated flood-attributable percent change in rate of hospitalization during the exposure period (lag 0) and each of the four following weeks (lags 1-4) in Figure \ref{level1_results}. We observed the largest flood-attributable increase in hospitalization rates for diseases of the musculoskeletal system and connective tissue in the fourth week after exposure (5.9 \%;  95 \% confidence interval [CI]: 4.0 \%, 7.8 \%). Risks of hospitalization due to diseases of the nervous system and sense organs were elevated during flood exposure and each of the four weeks after flooding, with a maximum in the third week following exposure (4.6 \%; 95 \% CI: 1.7 \%, 7.6 \%). Hospitalization admissions for diseases of the skin and subcutaneous tissue were similarly elevated during exposure and all four lag weeks, but with a peak during the flood exposure (5.1 \%; 95 \% CI: 1.9 \%, 8.4 \%). We also observed a consistently elevated risk for injury and poisoning across the exposure and lag periods, with a peak of 2.0 \% (95 \% CI: 0.4 \%, 3.7 \%). Conversely, we observed the largest flood-attributable decrease in hospitalization rates for respiratory diseases (-4.0 \%; 95 \% CI: -5.3 \%, -2.6 \%) during the first week after flood exposure. Diseases of the respiratory system were the only cause of hospitalization that demonstrated flood-attributable decreases across the exposure period and each lag week. 
\\ \\
We saw increases in hospitalization rates for mental illness during the flood exposure (3.0 \%; 95 \% CI: 0.4 \%, 5.6 \%) but no clear trends across the lag weeks. In contrast, there is an increased risk for circulatory system diseases and neoplasms (cancers) in lag weeks 3-4 following exposure. For diseases of the blood, digestive system, genitourinary system, and infectious and parasitic diseases, we did not observe noteworthy deviations from expected hospitalization rates across studied time periods. 
There was a decrease in endocrine system hospitalizations (-2.6 \%; 95 \% CI: -4.8 \%, -0.4 \%) during the second week after flood exposure. 
\\ \\
We also conducted a stratified assessment of the impacts of flood exposure by flood severity, i.e., to compare the health impacts of floods classified as moderate with those classified as either high or extreme severity. 
Across causes of hospitalization and lags, we generally found larger and more consistent adverse health impacts of high/extreme severity floods relative to those of moderate severity (Figure \ref{severity_results}). These exacerbated effects of more severe floods are especially pronounced for diseases of the nervous system and mental illness across the exposure period and all lag weeks, and for diseases of the skin during the exposure period and the first two lag weeks. For skin disease hospitalizations, we observed the highest estimates of a 7.9 \% (95 \% CI: 3.4 \%, 12.6 \%) increase 1 week after a high/extreme severity flood exposure compared with a non-significant decrease of -1.5 \% (95 \% CI: -5.9 \%, 3.2 \%) after exposure to floods classified as lower severity. For some disease types for which the overall effects of floods were null or reduced hospital visit rates, adverse effects were observed when restricting to high-severity floods. We found elevated risk of endocrine, metabolic, and immunity disorders during high-severity floods. Of note is the widening gap between hospitalization rates for respiratory diseases during and up to four weeks after flood exposure for moderate vs high/extreme severity floods. Namely, we saw a significant decrease (-8.0 \%; 95 \% CI: -9.8 \%, -6.3 \%) for floods of moderate severity in comparison to floods with high to extreme severity (3.5 \%; 95 \% CI: 1.4 \%, 5.7\%) at lag week 4. 
\\ \\
As a supplemental analysis, we examined whether there a difference exists in estimated effects of flood exposure and cause-specific hospitalization rates by type of hospital admission (emergency vs non-emergency) with a stratified analysis (Appendix \ref{er}). 
\\ \\
To understand if and how flood effects differ in more and less marginalized communities, we also investigated the associations between flood exposure and hospitalization rate stratified by percent of ZIP Code residents identifying as Black (percent Black above/below the median, 1.7\%). Results are given in Figure \ref{blk_results}. For nervous system diseases, we observed exacerbated flood effects for communities with lower percentages of Black residents across most lag periods and especially, during flood exposure (8.4 \%; 95 \% CI: 3.6 \%, 13.4 \%). A similar but less pronounced pattern emerged for endocrine system disorders (2.4 \%; 95 \% CI: -1.5 \%, 6.5 \%). These communities also had larger estimated effects during later lag periods for diseases of the musculoskeletal system and connective tissues, with a peak during the fourth week after exposure (8.2 \%; 95 \% CI: 5.1 \%, 11.5 \%). On the other hand, we observed larger flood-attributable increases in hospitalization for both infectious diseases and skin diseases for ZIP Codes with higher proportions of Black residents, compared to those with lower, during the exposure period and most lag weeks. ZIP Codes with higher percent Black residents also had a significant increase in hospitalizations for mental illness during flood exposure, while ZIP Codes with lower percent Black did not. 

\subsection{Sensitivity analyses}
In sensitivity analyses that considered different flood exposure definitions and that additionally included floods caused by snowmelt/ice or dam breakages, we found similar results, with only minor changes in magnitude (Appendix \ref{sensitivity}). 


\section{Discussion}\label{sec3}

In our study, we utilized nationwide Medicare data for a 17-year period and a collection of satellite-based flood maps to conduct the first large-scale study of how flood exposure affects the rate of hospitalizations in older adults for 13 critical disease causes. We observed increased hospitalization rates due to flood exposure for numerous disease categories, such as diseases of the musculoskeletal system, nervous system, and skin, in addition to increases in injury and mental illness-related hospitalizations. Other causes showed variable changes throughout and in the weeks after flood exposure. Exposure to higher severity floods amplified adverse impacts compared to lower severity flood events. The effects of floods on hospitalization rates revealed differing patterns in populations with higher vs lower proportions of Black residents. 
\\ \\
Previous work has investigated the impact of tropical cyclone exposure (defined based on location of origin of the storm and windspeed) on cause-specific hospitalization rates \citep{parks2021tropical} in the Medicare population. In contrast, our work focuses specifically on exposure to major flood events, most of which originate from non-tropical extreme precipitation events in the United States. Relative to prior work on tropical cyclones, we also considered longer post-exposure time periods (4 weeks in comparison to 1 week) because we hypothesized that the health impacts of flood exposure may materialize over a longer period of time compared to the impacts from exposure to high winds. 
In general, our findings differed considerably from studies on tropical cyclones and cause-specific hospitalization. While tropical cyclones have consistently led to increases in respiratory hospitalizations \citep{parks2021tropical,yan2021tropical}, we found evidence of decreases in respiratory hospitalization rates during and after flood exposure. Meanwhile, we found more compelling evidence that floods cause an increase in nervous system, musculoskeletal system, and mental illness-related hospitalizations, relative to tropical storms \citep{parks2021tropical}. For causes that shared similar trends for both floods and tropical storms, such as injuries and skin diseases, the magnitude of the estimated effects tended to be smaller for floods. 



\subsection{Nervous system diseases}
Diseases of the nervous system and sense organs include meningitis, encephalitis, Parkinson's disease, multiple sclerosis, paralysis, epilepsy, headaches, eye-related hospitalizations, vertigo and ear disorders.
Acute, provoked seizures and long-term post-traumatic epilepsy can arise from traumatic brain injuries due to the magnitude of force released by storm events \citep{sisodiya2019climate}. For individuals already diagnosed with epilepsy, flood events can induce stress, fatigue, and sleep deprivation which might result in a lack of seizure control and non-seizure epileptic aspects \citep{gulcebi2021climate}. Contaminated floodwater can induce headaches and eye irritation like conjunctivitis \citep{minovi_toxic_2020, ramesh2023adverse, saulnier2017no}. Otitis media can occur as a result of floodwater entering the inner chamber of the ear or other anatomical parts \citep{morgan_revisiting_2005, world2005flooding}.

\subsection{Skin and subcutaneous tissue diseases}
Elevated rates of skin and subcutaneous tissue diseases during and after flood exposure are plausible as floods can result in polluted and contaminated water sources, crowded shelters, and lack of sanitation \citep{centers2005infectious, de2016gastrointestinal}. The formation of standing or slow-moving water can lead to extended immersion in floodwater which poses risk to bodily extremities \citep{bandino2015infectious}. Together, this could result in cutaneous and fungal infections, infections of traumatic wounds, and insect bites \citep{tempark2013flood,dayrit2018impact}. Literature delineates the impact (moment of impact until day 4) and post-impact (day 4 up to day 28) phase of a flood, which correspond closely to the studied lag periods and represent the critical periods where dermatological implications are of pressing risk \cite{parker_skin_2022}. Atopic dermatitis, a chronic inflammatory skin disease, affecting both children and adults is associated with increased emergency visits after flood events \citep{kam_derm_mental_2023}. Furthermore, "cutaneous infections, immersion injuries, noninfectious contact exposures, and exacerbation of underlying skin diseases" have been reported after global catastrophic flood disasters including tsunamis in Asia (2004), Hurricane Katrina in United States (2005), and widespread floods in Pakistan (2010) where 700,000+ skin-related cases were documented \citep{parker_skin_2022, bandino2015infectious, centers2005infectious}. 

\subsection{Injury and poisoning}
There are several feasible explanations for increases in hospitalization rates for injuries and poisoning. In particular, injuries might occur from trying to evacuate or move personal belongings. Following the flood, injuries can arise from attempts to return to normalcy byways of decontaminating properties, demolition, and repairing structures \citep{ahern_global_2005, ramesh2023adverse}.
In high-income countries such as the United States, one risk factor is also the operation of vehicles into floodwater \citep{jonkman2005analysis}. Indirectly, injuries can result from damaged and unmaintained infrastructure such as was the case following Hurricane Katrina \citep{comfort2006cities}. From previous floods in the Midwest, reported hospitalizations largely included sprains and strains, contusions, and lacerations \citep{ahern_global_2005}. Other mechanisms for injury-related hospitalization include burns, electrocutions, and carbon monoxide poisoning due to petrol-powered electric generators and the use of pressure washers with poor indoor ventilation \citep{centers2000morbidity}. In general, following Hurricane Harvey and other flood events, studies observed increased emergency department visits related to injuries \citep{ramesh2023adverse}.

\subsection{Musculoskeletal system diseases}
Diseases of the musculoskeletal system include arthritis and other joint disorders, spondylosis, acquired deformities, lupus, and other connective tissue or bone diseases. An initial decrease in musculoskeletal system hospitalizations during floods is plausible because a large portion of these hospitalizations are for chronic sub-causes. For chronic conditions, individuals may delay care during the immediate flood exposure period due to risk of additional harm from traveling to a hospital during floods \citep{radcliff2018model}. Second, a primary function of the musculoskeletal system is to provide the ability to move. 
Increases in hospitalization rates for musculoskeletal system diseases in the weeks following flood exposure are plausible due to patients visiting the hospital for planned, non-emergent care that was disrupted due to the flood event or rescheduled \citep{radcliff2018model}. Additionally, flood clean-up often requires repetitive, unfamiliar movements such as bending and lifting heavy weights for long hours, forcing patients to withstand heavy strain on muscles and joints in their back, arms, and legs \citep{sihawong2012incidence, ohl2000flooding}. This can aggravate symptoms for existing conditions and drive increases in hospitalizations for non-emergent admissions. 

\subsection{Mental health-related impacts}
Flood events can have both direct and indirect effects on mental health. Direct impacts are well-documented for acute-stress disorders that can develop into post-traumatic stress disorder over time, anxiety, depression, and increase in substance use \citep{fernandez_mental_2015, ahern_global_2005, ohl2000flooding}. Indirect impacts arise from the intertwined relationship between physical and mental well-being in addition to changes in the natural and built environment. The sudden transitions and jarring shifts in physical conditions and the natural environment can lead to mood disorders \citep{fullilove1996psychiatric, kam_derm_mental_2023}. Potential pathways for impacts that contribute to increased anxiety and depression include property loss/damage and displacement, limited access to resources, death of loved ones, insurance problems, and loss of identity from social changes \citep{ahern_global_2005, fernandez_mental_2015}. The physical and mental toll of rebuilding following a flood disaster can also contribute to chronic stress. Furthermore, physical conditions such as atopic dermatitis are associated with increased risks of depression, anxiety, cognitive disorders, suicidal thoughts, adjustment and personality disorders, and alcohol use \citep{kam_derm_mental_2023}. This briefly demonstrates that physical health is influenced by mental health in the context of climate-related hazards.

\subsection{Respiratory system diseases}
We found an overall protective effect of flood exposure on respiratory hospitalizations, in contrast to studies of hurricanes which have found strong adverse burdens on short-term respiratory outcomes \citep{parks2021tropical,yan2021tropical,nethery2021integrated}. However, these negative impacts were largely hypothesized to be driven by power outages caused by strong winds (which disrupts use of electric-powered breathing equipment for people with COPD and other severe lung diseases), and we note that heavy precipitation events that lead to flooding are not always accompanied by high winds. Furthermore, in analyses stratified by flood severity, we found that the inverse associations between floods and respiratory hospitalizations were driven by less severe floods. For floods of high/extreme severity, estimated effects were null for lags 0-3 while a significant increase in hospitalizations was observed during lag week 4. Less severe floods may result in minimal flooding of residential areas, and the associated heavy precipitation could be protective for respiratory outcomes because rainfall reduces particulate matter concentrations and settles dust \citep{zhang2018influences}, both of which are respiratory irritants. The adverse effects of high severity floods, which generally flood more residential areas, on respiratory outcomes at lag week 4 could be due to indoor mold exposure and dampness in flooded homes.

\subsection{Other diseases}
Flood-attributable increases in cancer-related admissions in lag week 4, which were especially pronounced for high severity floods, could have been driven by flood-related cancer treatment disruptions, which could result in worse disease outcomes over the medium-to-long term \citep{man2018effect}. We also found evidence of adverse cardiovascular effects of floods only in later lag weeks, which could be a result of prolonged stress and physical exertion during cleanup \citep{steptoe2012stress}.
\\ \\
We observed little or no effect of flooding on hospitalization rates for infectious and parasitic diseases, digestive system diseases, or blood diseases. Previous literature has suggested that flooding events, which often result in stagnant water, can lead to the spread of waterborne diseases and a subsequent increase in infectious/parasitic, blood, and/or digestive system diseases \citep{ahern_global_2005, centers2005infectious, de2016gastrointestinal, howard1996infectious}. Increased gastroenteritis (digestive), outbreaks of leptospirosis (blood) and diarrhea (genitourinary) can arise from damage to sewage systems and poor sanitation and hygiene due to human displacement \citep{wade2014flooding, saulnier2017no, lowe_factors_2013}; however, a majority of the empirical evidence of these findings has been found in low-income countries while assessing less severe outcomes than inpatient hospitalization \citep{alderman2012floods}. In our high-income country setting, these effects are likely less prevalent and when present, they are likely to be primarily observable in outpatient and emergency department visit data. 
Similarly, while flood-related adverse impacts on the endocrine system are plausible, due to disruption of dialysis \citep{kelman2015dialysis} or diabetes treatment \citep{saulnier2017no}, we did not observe consistent evidence of effects on endocrine system hospitalizations here.

\subsection{Inequities and health impacts of flooding}
The consequences of stand-alone and repeated exposure to flood events are not borne equally by populations in the United States \citep{selsor_inequity_2023}. Certain subgroups of the population have been identified as socially vulnerable including racial minorities, children and seniors, and those with lower income or levels of education. Many of these disadvantaged groups disproportionately reside in areas susceptible to flooding \citep{bigi_vulnerability_2021, debbage_injustice_2019, qiang_disparities_2019} while having minimal flood response support and limited access to disaster relief services \citep{rufat_vulnerability_2015}. Furthermore, these pervasive inequalities are predicted to persist and possibly widen in coming years due to climate change \citep{coninx_integrating_2021, marshall_african_2015, wing_inequitable_2022}. This is particularly of note for the Southeastern United States, a region we identified as frequently experiencing flood events in \nameref{sec2}, where systemic racism and segregation have led to increased flood exposure for Black communities \citep{cutter_vulnerability_2003, ueland2006racialized}. While recent work by \citep{wing_inequitable_2022} reveals that current flood exposure risks are greatest in poor White communities in the United States, our results indicate that for certain health outcomes like skin diseases and mental illness, flood exposure had a larger impact in communities with a higher proportion of Black residents. In contrast, communities with lower proportions of Black residents experienced larger increases in nervous system diseases during flood exposure and musculoskeletal system diseases at later lags. Work by \citep{ramesh2023adverse} found that associations between flooding and concentration problems, which could be easily misdiagnosed, were lower among Black and African-American individuals as compared to other racial groups. 
\\ \\
While it is plausible that certain health impacts of flooding are borne more acutely by marginalized groups, another possible explanations for these differences are differences in access to care and implicit biases in how disease conditions are coded between racial groups. For instance, these factors may play a role in our findings that nervous system hospitalizations increase in communities with smaller Black populations during floods but do not increase significantly in communities with larger Black populations. Meanwhile, hospitalizations coded as mental illness-related increase during floods in communities with larger Black populations but not in communities with smaller Black populations. This could indicate differences in access to neurologic care across communities and/or differences in how the same conditions are coded for marginalized groups, e.g., the same condition, such as delirium, could be coded as mental illness for some groups and nervous system diseases for others. Previous research has found that both Black and non-Black individuals experience stress-related mental health issues at similar rates but Black individuals have less access to neurologic care \citep{robbins_black_2022, saadi2017racial}. Additionally, the Southeastern United States, where approximately 50 \% of Black people reside, has notably less outpatient neurologists per capita than the Northeast \citep{bradley2000neurology}. Flood attributable increases in skin disease hospitalizations were also more prominent in communities with larger Black populations. Minority populations have been found to experience greater incidence of skin disease with less access to healthcare \citep{buster2012dermatologic}, which may be exacerbated by climate change driven natural disasters.

\subsection{Strengths and limitations}
Our study is a comprehensive assessment of the impact of flood exposure on cause-specific hospitalization rates in older adults in the United States. We link Medicare hospitalization records to a multi-year flood exposure dataset for 13 major negative health outcomes and corresponding sub-causes across 46 states and districts in the United States. While our work is novel in explicitly quantifying the health implications of flood events for a vulnerable population, our study has limitations. 
First, while flooding tends to be a highly localized event, we summarized flood exposures at the ZIP Code level for linkage with Medicare records. Thus, not everyone living in an exposed ZIP Code necessarily experienced residential flooding. However, it is possible that flood events impact the health of even community members whose homes were not flooded and our approach allows us to capture any such effects. Moreover, we found that our results were robust to different exposure definitions. 
\\ \\ 
As in any observational study, residual confounding cannot be entirely ruled out. However, we leveraged a matched study design that controls for time-invariant factors across ZIP Codes and seasonality while additionally adjusting for longer-term trends in time and meteorological variables. Any remaining confounder would therefore have to covary with both hospitalization rate and flood exposure within ZIP Codes and be independent of the variables included in our model. 
Our design identifies two controls for each flood event by matching on ZIP Code and day-of-year in years prior or following exposure. In doing so, we do not take into account differences in hospitalization patterns between the weekdays vs weekends.
Including this information is non-informative when recognizing that floods often last multiple days (and can span up to 6 weeks), and sources of flooding do not systematically differ between workdays and weekends, so 
it is unlikely that the results we observe are attributable to this factor.

\section{Conclusion}\label{sec4}

Our work provides the first comprehensive assessment of how flooding influences cause-specific hospitalizations among older adults, and our findings can be utilized to allocate funding and resources towards flood preparedness and response. To better assist socially vulnerable populations, the construction of ample hygiene-forward shelters and the development of mobile health units that contain common medications will be crucial to improving human health in the aftermath of flood events. Our work identifies health-related flood implications that can be leveraged to improve hospital and physician preparation with the potential to save critical lives during flood disasters, a pressing public health concern. 

\section{Methods}\label{sec5}

\subsection{Study population}
We used Medicare inpatient claims data from the Centers for Medicare and Medicaid Service (CMS). We obtained data from Medicare beneficiaries that were aged 65 or older and enrolled in the fee-for-service program for at least 1 month from January 1, 2000 to December 31, 2016 and living in the contiguous United States \citep{cms}. From each billing claim, we extracted the date of admission, primary ICD-9-CM code (ICD-10-CM code for hospitalizations on or after October 1, 2015), and enrollee's ZIP Code of residence. Hospitalization counts for a given ZIP Code were computed by summing the total number of hospitalizations with a primary diagnosis for a specific disease. For the primary analyses, these counts include both emergency and non-emergency hospitalizations. Analyses were restricted to ZIP Codes that experienced at least one flood exposure during the study period, as defined below. 

\subsection{Outcome assessment}
We obtained the primary diagnosis for each hospitalization among Medicare Fee For Service enrollees residing in each of the 11,594 United States ZIP Codes that experienced at least one severe flood event between 2000 and 2016. During the flooded and control dates identified for each ZIP Code, there were 4,879,643 recorded hospitalizations. We grouped 13,726 International Classification of Diseases, Ninth Revision, Clinical Modification (ICD-9-CM) codes and 72,446 International Classification of Diseases, Tenth Revision, Clinical Modification (ICD-10-CM) codes using the Clinical Classifications Software (CCS) algorithm to 18 mutually-exclusive, clinically meaningful CCS level 1 causes of hospitalization \citep{elixhauser2014clinical}. The CCS algorithm was developed by the Agency for Healthcare Research and Quality and was initially created for ICD-9-CM codes. Mappings to ICD-10-CM codes were later developed as a Beta release of the CCSR algorithm for ICD-10-CDM codes \citep{castro_2020}. Out of the 18 disease causes, we exclude five that are ill-defined or not relevant to the Medicare cohort such as those related to pregnancy or infertility. This leaves the following 13 level 1 causes: infectious and parasitic diseases, neoplasms, endocrine system disorders, blood diseases, nervous system diseases, circulatory system diseases, respiratory system diseases, digestive system diseases, genitourinary system diseases, skin and subcutaneous tissue diseases, musculoskeletal system diseases, injury and poisoning, and mental illness. 

\subsection{Flood exposure assessment}

We obtained data on severe flood exposures that occurred in the contiguous United States from 2000-2016 using the Global Flood Database, the most comprehensive collection of satellite-based high-resolution historic flood maps to date \citep{tellman_satellite_2021}. The creation of these flood maps was described in detail by \citep{tellman_satellite_2021}, and we summarize key points and features here. To construct the flood maps, dates and approximate locations of major flood events were identified using the Dartmouth Flood Observatory event catalogue \citep{dfo}. The Dartmouth Flood Observatory includes more flood events than other listings such as the Emergency Events Database (Em-Dat) \citep{emdat_2000}, identifies the proposed cause of the flood, and provides critical spatial information that can be used to leverage flood detection algorithms on satellite imagery. 
\\ \\
High-resolution flood maps were created for a selection of events in the Dartmouth Flood Observatory catalogue where satellite imagery was capable of producing inundation footprints with high accuracy. Flood maps were generated  by analyzing daily satellite imagery from NASA's Moderate Resolution Imaging Spectroradiometer (MODIS) satellites at 250-meter resolution over the relevant area and time period \citep{modis}. To do this, the following steps were taken. First, the authors defined the region of interest to map for the potential flood event by selecting all global HydroSHEDS watersheds that intersect with the Dartmouth Flood Observatory event polygon. 
Then, an existing algorithm was deployed on satellite-measured reflectance to classify each pixel in the region of interest as surface water or non-surface water. Quality control measures were undertaken after, e.g., to reduce misclassification from other weather elements such as cloud shadows, and accuracy was assessed.
Details on algorithms and other technical aspects are provided in \citep{tellman_satellite_2021}.  
\\ \\
The flood map for each flood event is a raster composed of 250 m$^2$ pixels. Each flood map consists of five bands \citep{tellman_satellite_2021}, three of which we describe here for use in subsequent analyses. 
\begin{itemize}
\item \textit{flooded}: a binary indicator for whether each pixel value was considered flooded during the event (i.e., whether the satellite detected surface water during the period of the given flood event) 
\item \textit{duration}: a discrete numeric variable for the number of days each pixel was assigned flooded during the event 
\item $\textit{jrc\_perm\_water}$: a binary indicator for whether each pixel value revealed the presence of surface water consistently for observations in 1984-1999 and 2000-2018 
\end{itemize} 
In order to isolate flooded pixels and distinguish them from permanent bodies of water, we masked permanent water from the flood bands using the \texttt{raster} package, version 3.6-14 in R, version 4.1.0 \citep{raster, R}. 
\\ \\
These maps were used to identify ZIP Codes and days where floods occurred in the contiguous United States. To link flood exposure information with our health records, we aggregated it to the ZIP Code level, which is the finest geographic resolution at which residential location is recorded in Medicare claims data. We computed percent of ZIP Code area flooded and the mean and maximum duration of flooding in flooded pixels within the ZIP Code using ZIP Code polygon shapefiles. Because ZIP Code boundaries can change over time, we used the corresponding shapefile for the year of the flood event or the closest year preceding the event for years in which no shapefile existed. We defined a particular day and ZIP Code as exposed to a flood if at least 0.5 \% of the ZIP Code's surface area was flooded on the given day or if the area flooded was at least 5 square-miles.  The flooded ZIP Codes identified for each flood event comprise a master dataset that we combined with information from the Dartmouth Flood Observatory catalogue such as a unique 4-digit flood identifier, severity, main cause, and start/end dates. These data are accessible at \citep{flood2023dataverse}. For our primary analysis, we only considered floods caused by heavy rain or tropical storms, as those caused by dams or snowmelt/ice melt tend to have longer durations and may not be naturally occurring, in the case of dams. 


\subsection{Covariate data}
We obtained temperature, humidity, and windspeed data from the Gridded Surface Meteorological (gridMET) Dataset \citep{gridmet}. GridMET integrates gridded climate data from the Parameter-elevation Regressions on Independent Slopes Model (PRISM) \citep{prism} with regionally reanalyzed data from NLDAS-2 \citep{nldas2_1,nldas2_2} to produce daily estimates of climatological properties at the Earth's surface across the contiguous United States. Validation is performed against a complex network of weather stations \citep{gridmet}. We aggregated gridded daily estimates of maximum air temperature, maximum relative humidity, and wind velocity at 10m at a resolution of 4-km x 4-km to the ZIP Code level using the \texttt{exactextractr} R package (version 0.9.1), taking into account the coverage fraction of each grid cell \citep{exactextractr}. 


\subsection{Study design}
For each flood event, we first determined each ZIP Code's exposure period (the set of days when that particular ZIP Code was flooded). Then, we matched each ZIP Code to two control periods from the same ZIP Code and the same time period (i.e. days-of-year) but in years preceding or following flood exposure. Two control periods were identified in order to adjust for bidirectional time trends. Matching time periods were selected on the basis of the closest two years with no flood exposure during the equivalent days-of-year or lag time (four subsequent weeks). These control periods allowed us to establish a baseline rate of hospitalization (absent flood) in each ZIP Code under study. Two control periods With this study design, comparisons between the flooded ZIP Codes and matched controls addresses both time-invariant ZIP Code features (by matching on ZIP Code) and seasonal trends in the outcome (by matching on day-of-year) that may otherwise bias results. We defined a ``stratum'' as a given ZIP Code-exposure period, its matched control periods, and all of the lag weeks following each flooded and control period. ZIP Code hospitalization counts were aggregated over each exposure/control period and each lag week separately. Our analytic sample was composed of the data for all strata across the 75 flood events. 
\\ \\
We formally assessed the change in rates of hospitalization relative to business-as-usual during the flood exposure and the subsequent four weeks via statistical modeling. We fit conditional quasi-Poisson models to each cause of hospitalization separately, modeling rates of hospitalization in flood-exposed ZIP Codes on exposure days, their matched control days, and in the four lag weeks following both exposures and controls. 
This approach accounts for potentially overdispersed outcomes and represents the case-crossover study design analogue for area/ecologic data. Lastly, the model included unconstrained distributed lag terms to examine how hospitalization rates deviate from expected rates during and up to four weeks after flood exposure. Four lag weeks were employed to investigate possibly delayed, medium-term health impacts of flood exposure.
\\ \\
We index strata by $s$ and each of the units/periods within strata by $t$. The model takes the following form: 
     
\begin{equation}\label{primary_model}
    \begin{split}
        \log \left( E[Y_{st}] \right) &= \alpha_0 + \alpha_{s} + \sum_{\ell=0}^4 \beta_{\ell} \text{Exposure}_{\ell s t} + \boldsymbol{\gamma}' \boldsymbol{z}_{st} + \log(\text{person-days}_{s t})
    \end{split}
\end{equation}

where $Y_{st}$ is the count of CCS level 1-defined cause-specific hospitalizations in period $t$ within stratum $s$; $\alpha_{s}$ are the stratum-specific intercepts; Exposure$_{lst}$ is an indicator of whether the given period is the exposure period ($\ell=0$) or one of the lag weeks following an exposure ($\ell\in [1,4]$); $\boldsymbol{z}_{st}$ is the vector of confounder values; and $log(\text{person-days}_{st})$ is the perscanon-time offset, computed as the number of Medicare enrollees in the corresponding ZIP Code and year multiplied by the length of the period in days. The $\beta_\ell$ are log hospitalization rate ratios for exposure vs control periods and their corresponding lags. The $\boldsymbol{\gamma}$ are the regression coefficients corresponding to the vector of confounders. 
\\ \\ 
We report estimated flood-attributable percent changes in hospitalization rates for the exposure period and each lag as $100\times\left(\text{exp}(\widehat{\beta}_\ell) - 1\right)$. We define $\sum_\ell \left(\text{exp}(\widehat{\beta}_\ell) - 1\right)$ to be the cumulative effect of flood exposure on hospitalizations over the duration of the flood and four lag weeks. 
\\ \\ 
We adjusted for potential confounders that were not accounted for through matching, such as long-term trends in hospitalization patterns and day-to-day metereology by including natural splines with four degrees of freedom for year and averages of air temperature, relative humidity, and windspeed. To address multiple comparisons, we utilized the Bonferroni-Holm method to correct 95 $\%$ confidence intervals in our primary analysis by using $\alpha = 0.05/13$, since we consider 13 CCS level 1 causes. 
\\ \\
We investigated whether estimated effects varied by flood severity using the same model described previously. As defined by the Dartmouth Flood Observatory and adopted by the Global Flood Database, severity class takes on 3 distinct values: 1, 1.5, or 2 which we describe as ``moderate'', ``high'' and ``extreme'', respectively \citep{dfo}. Severity class is determined based on the scale of the event and expected recurrence intervals using historical data. Namely, floods of class moderate are large events with substantial damage/fatalities that have a 10-20 \% chance of occurring in a single year. Floods of class high are larger with recurrence interval of 21-99 years and/or locally recurring within 1-2 decades and affecting a significantly large geographic region. Floods of class extreme are the most severe, with recurrence greater than 10 decades \citep{dfo}. We compared floods of less severity ($\leq 1$) with those of high/extreme severity ($> 1$). 
\\ \\
As a secondary analysis, we fit stratified models to examine how associations between flood exposure and CCS level 1 cause-specific hospitalizations differ for marginalized communities, specifically communities with a higher proportion of Black residents at the ZIP Code level. Because ZIP Codes cover smaller surface area than other spatial units such as counties or states, they tend to be relatively homogeneous which allows for an informative analysis regarding the impacts of flood exposure on marginalized populations. We linked demographic data from the United States Census and the American Community Surveys to the aggregated Medicare hospitalization records as described previously by ZIP Code and year \citep{wu2020evaluating}. We removed 56 ZIP Codes for which demographic data were unavailable. Then, for each outcome, we fit separate models for ZIP Codes where the proportion of Black residents was less/greater than the $50^{th}$ percentile (median). 

\subsection{Sensitivity analyses}
We additionally assessed the sensitivity of our results to the exposure criteria we established. We fit models with all ZIP Codes that experienced flood exposure during the study period, regardless of percent and/or area flooded as well as only ZIP Codes that satisfied a more stringent inclusion criteria (at least 1 $\%$ flooded and/or had a flooded area of at least 10 square-miles). Lastly, we fit a model that includes ZIP Codes impacted by floods caused by dam faults or breakage and snowmelt/ice in addition to the causes considered in our primary analysis (heavy rain and tropical storms). Our results were robust to the sensitivity analyses described above and are shown in Appendix \ref{sensitivity}. 

\backmatter

\bmhead{Supplementary information}

\bmhead{Acknowledgments}

This work was supported by the Harvard Data Science Initiative, Harvard Graduate Prize Fellowship, the Sloan Foundation grant G20201394, and by NIH grants K01ES032458, R00ES033742, T32ES007142, and T32ES7069. The computations in this paper were run on the FASRC Cannon cluster and FAS Secure Environment supported by the Faculty of Arts and Sciences Division of Science Research Computing Group at Harvard University. 

\section*{Declarations}


\bmhead{Conflicts of interest}
The authors have no conflicts of interest to disclose.

\bmhead{Ethics approval}
This study was approved by the Harvard Longwood Campus Institutional Review Board, IRB20-1910 Tropical Cyclones. 

\bmhead{Availability of data and materials}
Flood data is available via Harvard Dataverse \citep{flood2023dataverse}. Medicare enrollee data are publicly accessible, upon purchase after an application process, from the CMS \citep{cms}. 

\bmhead{Code availability}
Code for analysis and visualizations presented in this manuscript is available at www.github.com/NSAPH-Projects/floods-hospitalizations-glm. 

\bmhead{Author contributions}
S.A and R.M.P. collated and organized hospitalization files. S.A collated and constructed flood datasets from the database created by J.A.S and all authors developed the statistical model, which was implemented by S.A. S.A performed the analysis, with input from R.C.N and J.K.H who also contributed to study concept and interpretation of results. S.A. and R.C.N. wrote the first draft of the paper; all authors contributed to revising and finalizing the paper.





\begin{appendices}










\section{Sensitivity analyses}\label{sensitivity}


\subsection{Floods caused by heavy rain, tropical storms, ice/snowmelt, or dam faults with any level of exposure}
\includegraphics[scale = 0.42]{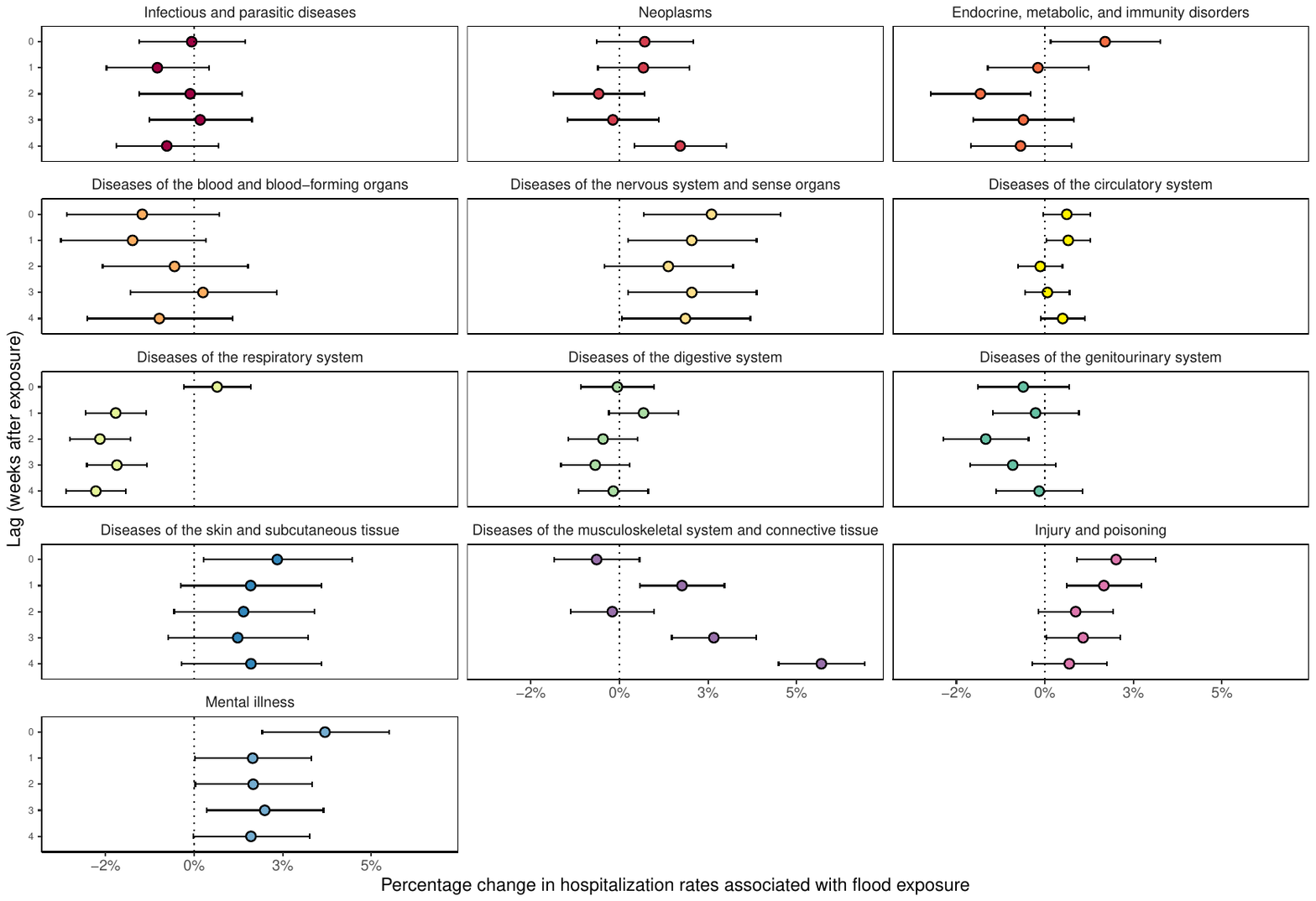}
\subsection{Floods caused by heavy rain or tropical storms with any level of exposure}
\includegraphics[scale = 0.42]{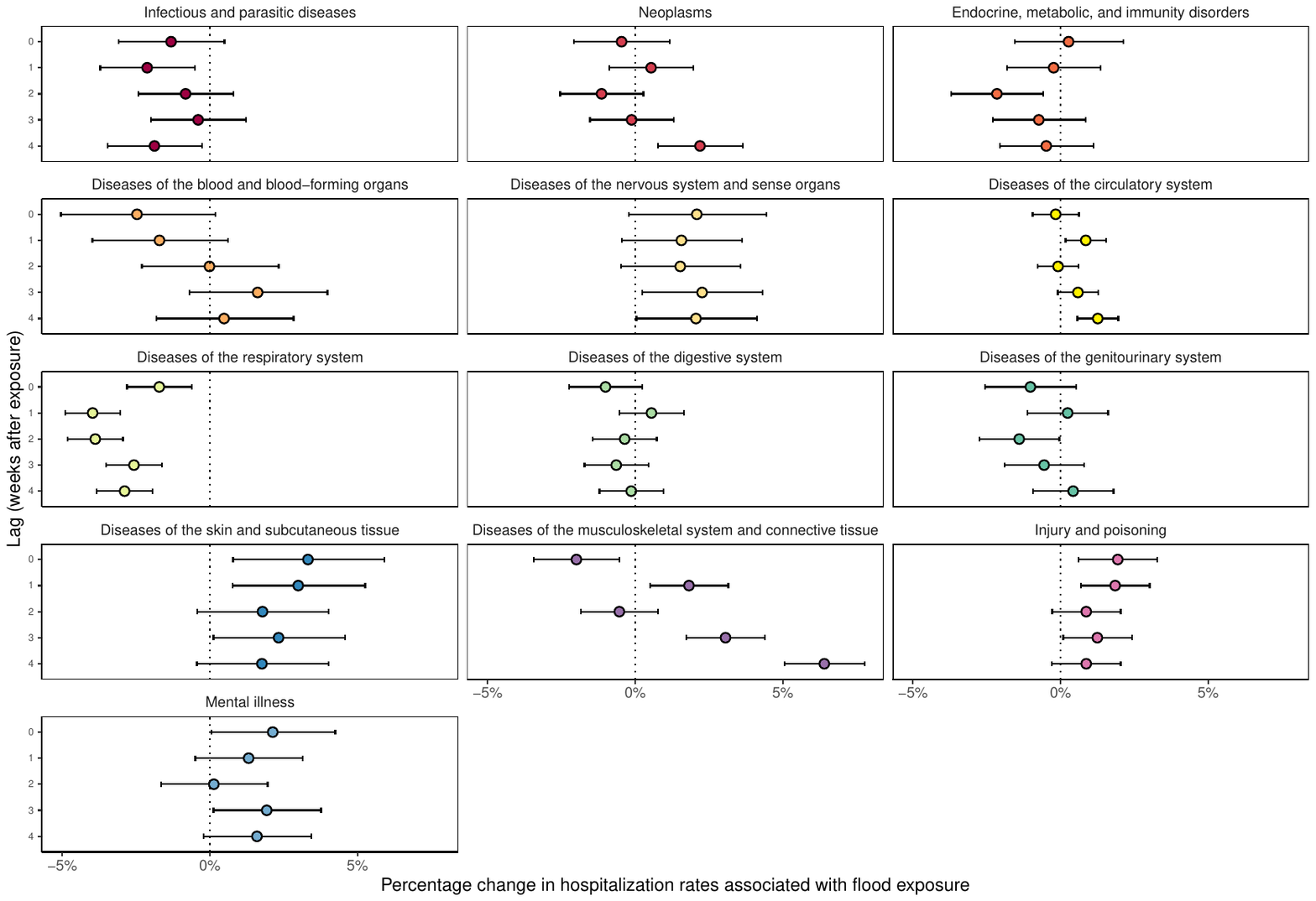}
\subsection{Floods caused by heavy rain or tropical storms with any level of exposure with at least 1 $\%$ flooded and/or a flooded area of at least 10 square-miles}
\includegraphics[scale = 0.42]{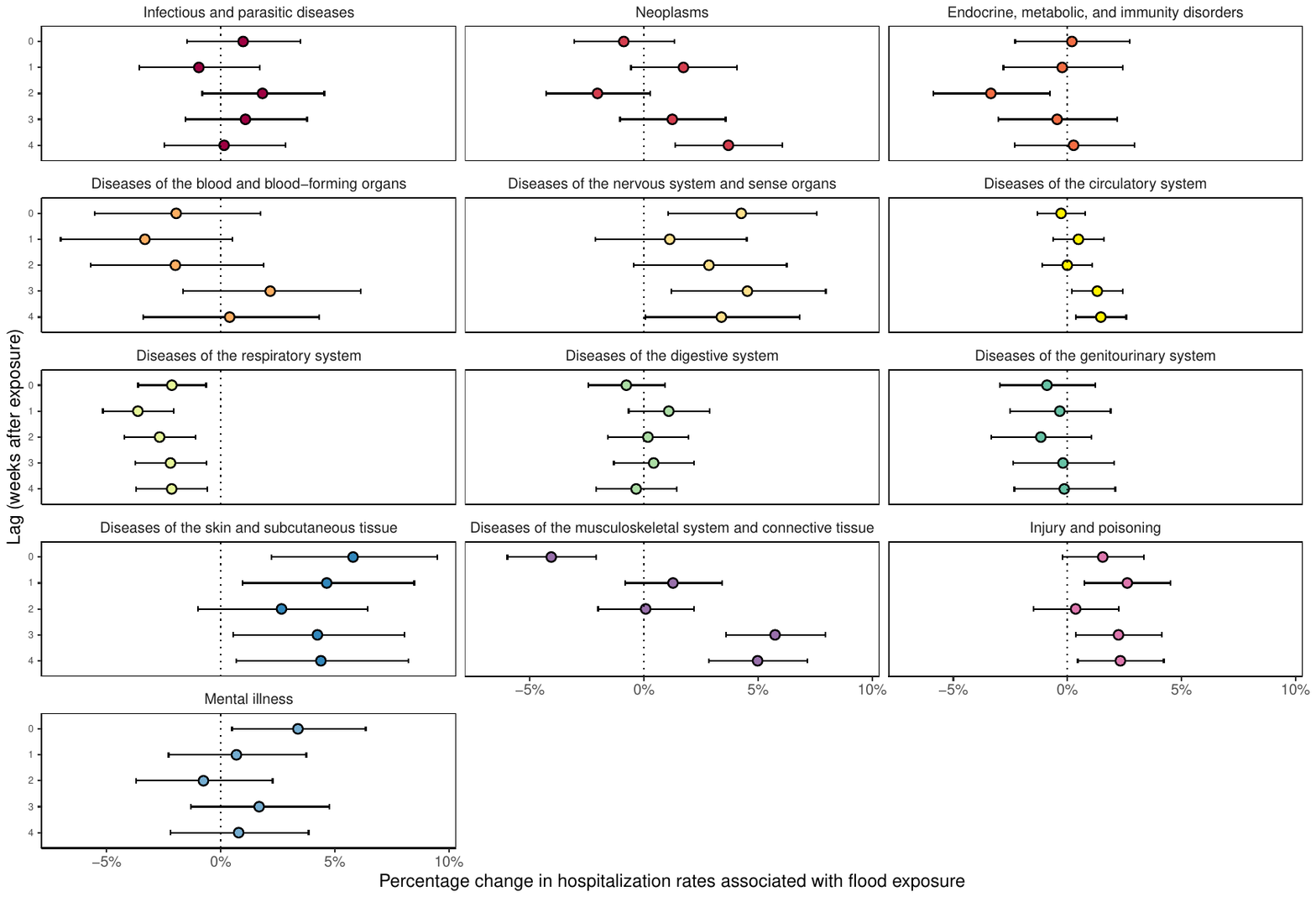}

\section{Additional analyses}\label{er}
\begin{figure}
    \centering
    \includegraphics[scale = 0.42]{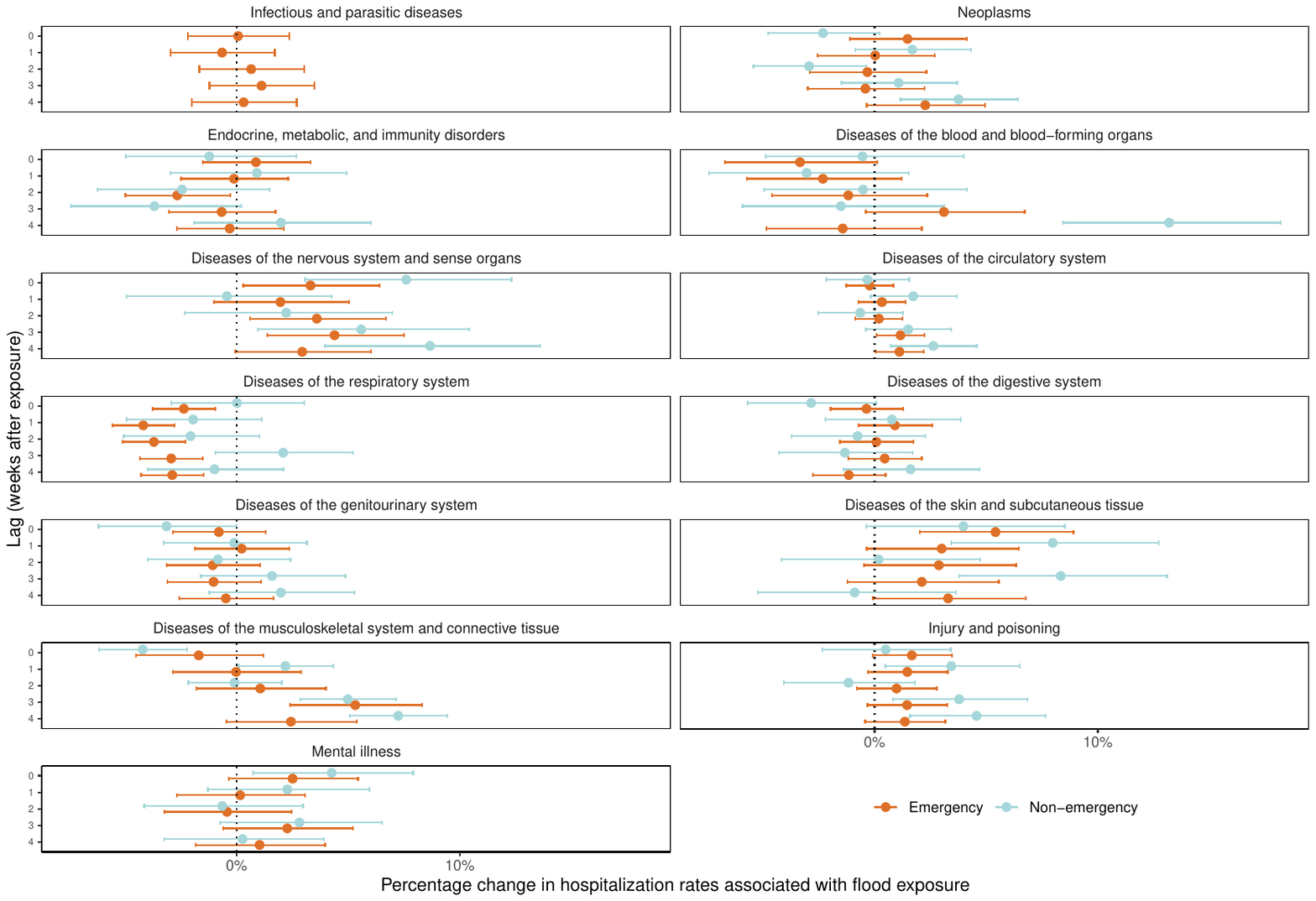}
    \caption{Percent change in cause-specific hospitalization rates with flood exposure by cause, type of hospital admission, and lag time}
    \vspace{128in}
    \label{er_results}
\end{figure}

We also examined the association between flood exposure and hospitalization rates by type of hospital admission (emergency vs non-emergency) as shown in Figure \ref{er_results}. Generally, non-emergency hospitalization rates decreased or showed no significant change during the flood (lag 0) except for nervous system diseases, mental illness, and skin and subcutaneous tissue diseases. For the aforementioned causes, we saw increases in non-emergency hospitalizations for multiple lag weeks. Emergency hospitalizations increased for the exposure period and all lag weeks for nervous system diseases, injury-related hospitalizations, and skin and subcutaneous tissue diseases. Conversely, for respiratory system diseases, we observed decreases across all lag periods for emergency hospitalizations and non-emergency hospitalizations (except lag 3 for the latter). For other causes, there tended to be little to no changes for emergency hospitalizations. Due to low case counts for infectious and parasitic non-emergency hospitalizations, we were unable to successfully fit a model that adjusted for potential meteorological confounders.

\end{appendices}


\end{document}